\pdfoutput=1 

\documentclass[10pt, conference, compsocconf]{IEEEtran}

\usepackage[pass]{geometry}
\ifdefined\RELEASE
\else
\usepackage{fancyhdr}
\fi

\usepackage[style=numeric,backend=bibtex,minnames=2,maxnames=20,sorting=ydnt,giveninits=true,sortcites,backref=false]{biblatex}
\usepackage{hyphenat}
\usepackage{multirow}
\usepackage{numprint}
\usepackage{graphicx}
\usepackage[normalem]{ulem}
\usepackage{xspace}
\usepackage[caption=false,font=footnotesize]{subfig}
\usepackage{url}
\usepackage{transparent}
\newcommand\stitchref[2]{\ref{#1}\subref{#2}}
\usepackage{tabularx}
\usepackage[table,dvipsnames]{xcolor}
\usepackage{microtype}

\emergencystretch=\maxdimen
\newcommand{\squishlistnum}{  %
 \newcounter{qcounter}
 \begin{list}{\roman{qcounter})~}{\usecounter{qcounter}}
  { \setlength{\itemsep}{0pt}
     \setlength{\parsep}{0pt}
     \setlength{\topsep}{0pt}
     \setlength{\partopsep}{0pt}
     \setlength{\leftmargin}{0em}
     \setlength{\labelwidth}{0em}
     \setlength{\labelsep}{0em} } }

\newcommand{\squishlist}{ %
 \begin{list}{$\bullet$}
  { \setlength{\itemsep}{2pt}
     \setlength{\parsep}{0pt}
     \setlength{\topsep}{2pt}
     \setlength{\partopsep}{0pt}
     \setlength{\leftmargin}{1em}
     \setlength{\labelwidth}{1em}
     \setlength{\labelsep}{0.5em} } }

\newcommand{\squishend}{
  \end{list}  }

\definecolor{amber}{rgb}{1.0, 0.49, 0.0}
\definecolor{amethyst}{rgb}{0.6, 0.4, 0.8}

\newcommand\sam{GRASP\xspace}
\newcommand\samnospace{GRASP\xspace}
\newcommand{\samacronym}{{\em GRASP -- \underline{GRA}ph-\underline{SP}ecialized\xspace}}

\newcommand\DBCM{GRASP\xspace}
\newcommand\noindentsectiontitle[1]{\vspace{0.3\baselineskip}\noindent {\em #1}}
\newcommand\noindentsubsectiontitle[1]{\vspace{0.15\baselineskip}\noindent {\em #1}}

\newcommand\visiblespace{\vspace{0.3\baselineskip}}

\newcommand\reducetablecaptionspace{\vspace*{-0.75\baselineskip}}
\newcommand\reducefigcaptionspace{\vspace*{-1.5\baselineskip}}
\usepackage{tikz}
\newcommand*\smalltextcircled[1]{\tikz[baseline=(char.base)]{
    \node[shape=circle,draw,inner sep=1pt] (char) {\small #1};}}

\newcommand\tw{\emph{tw}}
\newcommand\fr{\emph{fr}}

\newcommand\kr{\emph{kr}}
\newcommand\lj{\emph{lj}}
\newcommand\pld{\emph{pl}}
\newcommand\sd{\emph{sd}}
\newcommand\uni{\emph{uni}}

\newcommand\pr{PR}
\newcommand\prd{PRD}
\newcommand\sssp{SSSP}
\newcommand\bc{BC}
\newcommand\radii{Radii}

\newcommand\gorder{Gorder}
\newcommand\ie{{i.e., }}
\newcommand\eg{{e.g., }}
\newcommand\etal{et al. }

\newcommand\num[1]{\footnotesize{\numprint{#1}}}
\npthousandsep{,}\npthousandthpartsep{}\npdecimalsign{.}

\usepackage{hyperref}
\hypersetup{
    colorlinks=true,
    filecolor=magenta,    
    bookmarks=true,
    breaklinks=true,
    colorlinks=true,
    linkcolor=blue,
    citecolor=ForestGreen,
    urlcolor=cyan,
    pdfauthor={Priyank Faldu et al.},
    pdftitle={Domain-Specialized Cache Management for Graph Analytics}
}

\begin{document}
\sloppy
\title{Domain-Specialized Cache Management for Graph Analytics}
\author{
\IEEEauthorblockN{Priyank Faldu}
\IEEEauthorblockA{The University of Edinburgh \\ priyank.faldu@ed.ac.uk}
\and
\IEEEauthorblockN{Jeff Diamond}
\IEEEauthorblockA{Oracle Labs \\ jeff.diamond@oracle.com}
\and
\IEEEauthorblockN{Boris Grot}
\IEEEauthorblockA{The University of Edinburgh\\ boris.grot@ed.ac.uk}
}

\ifdefined\RELEASE
\else
\fancypagestyle{firstpage}{
    \fancyhf{}
    \renewcommand{\headrulewidth}{0pt}
    \fancyhead[C]{{\textbf{In Proceedings of the 26\textsuperscript{th} International Symposium on High-Performance Computer Architecture (HPCA'20)}}}
    \pagenumbering{arabic}
}
\fi

\maketitle

\ifdefined\RELEASE
\else
\thispagestyle{firstpage}
\pagestyle{plain}
\fi

\begin{abstract}

Graph analytics power a range of applications in areas as diverse as finance, networking and business logistics. A common property of graphs used in the domain of graph analytics is a \emph{power-law distribution} of vertex connectivity, wherein a small number of vertices are responsible for a high fraction of all connections in the graph. These richly-connected, \emph{hot}, vertices inherently exhibit high reuse. However, this work finds that state-of-the-art hardware cache management schemes struggle in capitalizing on their reuse due to highly irregular access patterns of graph analytics.

In response, we propose GRASP, domain-specialized cache management at the last-level cache for graph analytics. GRASP augments existing cache policies to maximize reuse of hot vertices by protecting them against cache thrashing, while maintaining sufficient flexibility to capture the reuse of other vertices as needed. GRASP keeps hardware cost negligible by leveraging lightweight software support to pinpoint hot vertices, thus eliding the need for storage-intensive prediction mechanisms employed by state-of-the-art cache management schemes. On a set of diverse graph-analytic applications with large high-skew graph datasets, GRASP outperforms prior domain-agnostic schemes on all datapoints, yielding an average speed-up of 4.2\% (max 9.4\%) over the best-performing prior scheme. GRASP remains robust on low-/no-skew datasets, whereas prior schemes consistently cause a~slowdown.

\end{abstract}


\section{Introduction}
\label{sec:intro}

Graph analytics is an exciting and rapidly growing field with applications spanning diverse areas such as optimizing routes, uncovering latent relationships, pinpointing \mbox{influencers} in social graphs, and many more. Graph analytics commonly process large graph datasets whose main memory footprint span tens to hundreds of gigabytes~\cite{twitter,pld}. When processing such large graphs, graph-analytic applications exhibit a lack of cache locality, leading to frequent misses in on-chip caches, compromising application performance~\cite{dbg,locality-exists,imp,prefetch-data-structure,gorder,fc,hubcluster}. %

A distinguishing property of graph datasets common in many graph-analytic applications is that the vertex degrees follow a skewed {\em power-law distribution}, in which a small fraction of vertices have many connections while the majority of vertices have relatively few connections~\cite{dbg,power-law,power-law-internet,powergraph,fc,hubcluster}.
Graphs characterized by such a distribution are known as {\em natural} or {\em scale-free} graphs and are prevalent in a variety of domains, including social networks, computer networks, financial networks, semantic networks, and airline networks.

We observe that the skew in the degree distribution means that a small set of vertices with  a large fraction of connections is responsible for a major share of off-chip memory accesses. The fact that these richly-connected vertices, {\em hot vertices}, comprise a small fraction of the overall footprint while exhibiting high reuse makes them prime candidates for caching. Meanwhile, the rest of the vertices, {\em cold vertices}, comprise a large fraction of the overall footprint while exhibiting low or no reuse.

We find that the existing caches struggle in exploiting the high reuse inherent in hot vertices due to the following two reasons: 
First, graph-analytic applications are notorious for exhibiting irregular access patterns that cause severe \emph{cache thrashing} when processing large graphs.
Accesses to a large number of cold vertices are responsible for thrashing, often forcing hot vertices out of the cache.
Second, hot vertices are sparsely distributed throughout the memory space, exhibiting a lack of spatial locality.  When hot vertices share the same cache block with cold vertices, valuable cache space is underutilized. 
Existing software techniques~\cite{dbg,fc,hubcluster} solve the latter problem by reordering vertices in the memory space, such that hot vertices share cache blocks with other hot vertices. 
However, the former problem of protecting hot vertices from premature evictions remains an open challenge, even for the state-of-the-art thrash-resistant hardware cache management schemes.

Almost all prior works on hardware cache management (\ie cache replacement) that target cache thrashing are {\em domain-agnostic}. 
These hardware schemes aim to perform two tasks: (1) identify cache blocks that are likely to exhibit high reuse, and (2) protect high reuse cache blocks from cache thrashing. To accomplish the first task, these schemes deploy either probabilistic or prediction-based hardware mechanisms
~\cite{dip,rrip,tadip,gippr,segmentedLRU,pipp,counter,sampler,ship,mdpp,hawkeye,harmony,perceptron,multi,leeway}. 
However, our work finds that graph-dependent irregular access patterns prevent these schemes from correctly learning which cache blocks to preserve, rendering them deficient for the broad domain of graph analytics. Meanwhile, to accomplish the second task, recent work proposes \emph{pinning} of high-reuse cache blocks in LLC to ensure that these blocks are not evicted~\cite{xmem}. However, we find that pinning-based schemes are overly rigid and result in sub-optimal utilization of cache~capacity.

\visiblespace

To overcome the limitations of existing hardware cache \mbox{management} schemes, we propose \samacronym Last-Level Cache (LLC) management. 
To the best of our knowledge, this is the first work to introduce domain-specialized cache management for the domain of graph analytics. 
\sam augments existing cache insertion and hit-promotion policies to provide preferential treatment to cache blocks containing hot vertices to shield them from thrashing.
To cater to the irregular access patterns, \sam policies are designed to be flexible to cache other blocks \mbox{exhibiting} reuse. 
Unlike pinning, \sam \mbox{maximizes} cache efficiency based on observed access patterns.

\sam relies on lightweight software support to \mbox{accurately} pinpoint hot vertices amidst irregular access patterns, in contrast to state-of-the-art schemes that rely on storage-intensive hardware mechanisms. By leveraging existing vertex reordering techniques, \sam enables a lightweight software-hardware interface comprising of only a few \mbox{configurable} registers, which are programmed by software using its knowledge of the graph data~structures.

\sam requires minimal changes to the existing microarchitecture as \sam only augments existing cache policies and its interface is lightweight.
\sam does not require additional metadata in the LLC or storage-intensive prediction tables. 
Thus, \sam can easily be integrated into commodity server processors, enabling domain-specific acceleration for graph analytics at minimal hardware cost. 

\visiblespace

To summarize, our contributions are as follows:

\squishlist

    \item We qualitatively and quantitatively show that a wide range of state-of-the-art domain-agnostic cache management schemes, despite their sophisticated prediction mechanisms, are inefficient for the domain of graph analytics.
    
    \item We introduce {\em \samnospace}, graph-specialized LLC management for graph analytics in processing natural graphs. \sam augments existing cache policies to protect hot vertices against cache thrashing while also maintaining flexibility to capture reuse in other cache blocks. \sam leverages a lightweight software interface to pinpoint hot vertices amidst irregular accesses, which eliminates the need for prediction metedata storage at the LLC, keeping the existing cache structure largely unchanged.
    
    \item Our evaluation on several multi-threaded graph-analytic applications operating on large, high-skew datasets shows that \sam outperforms state-of-the-art domain-agnostic schemes on all datapoints, yielding an average speed-up of 4.2\% (max 9.4\%) over the best-performing prior scheme.
    \sam is also robust on low-/no-skew datasets whereas prior schemes consistently cause a slowdown.
    
\squishend

\section{Motivation}
\label{motivation}

\begin{table}[!t]
    \caption{\footnotesize{Rows \#2 and \#4 show the percentage of vertices having degree equal or greater than the average (\ie hot vertices), with respect to in-edges and out-edges, respectively; the higher the skew, the lower the percentage.
    Rows \#3 and \#5 show the percentage of in-edges and out-edges connected to the hot vertices, respectively; the higher the skew, the higher the~percentage.
    }}
    \label{tab:hot-vertices-skew}
    \reducetablecaptionspace
    \centering
    \footnotesize
    \begin{tabular}{|c|c||r|r|r|r|r|}
    \hline
            & \centering{\bf Dataset}               & {\bf\lj}   & {\bf\pld}  & {\bf\tw}   & {\bf\kr}   & {\bf\sd}   \\ \hline \hline
        In  & Hot Vertices (\%)     & 25    & 16    & 12    & 9     & 11    \\  
        Edges & Edge Coverage (\%)  & 81    & 83    & 84    & 93    & 88    \\ \hline
        Out & Hot Vertices (\%)     & 26    & 13    & 10    & 9     & 13    \\
        Edges & Edge Coverage (\%)  & 82    & 88    & 83    & 93    & 88    \\ \hline
    \end{tabular}
\end{table}

\subsection{Skew in Natural Graphs}
\label{skew}
A distinguishing property of natural graphs is the \emph{skew} in their degree distribution~\cite{dbg,power-law,power-law-internet,powergraph,fc,hubcluster}. The skew follows a power-law, with the vast majority of the vertices having relatively few edges and a small fraction of vertices featuring a large number of edges. Such skewed distribution is prevalent in many domains and found, for instance, in nodes in large-scale communication networks (e.g., the internet), web pages in the web graph, and individuals in social graphs.

Table~\ref{tab:hot-vertices-skew} quantifies the skew for the datasets evaluated in this work (more details of the datasets in Table~\ref{tab:datasets}). For example, in the Twitter~\cite{twitter} dataset (labeled \tw), 12\% of total vertices are classified as hot vertices in terms of their in-degree (10\% for out-degree) distribution. These hot vertices are connected to 84\% of all in-edges (83\% of all out-edges) in the graph. Similarly, in other datasets, 9-26\% of vertices are classified as hot vertices, which are connected to 81-93\% of all edges.
In the following sections, we explain how this skew can be leveraged to improve cache~efficiency. %

\subsection{Graph Processing Basics}
\label{graph-processing-basics}
The majority of shared-memory graph frameworks are based on a vertex-centric model, in which an application computes some information for each vertex based on the properties of its neighbouring vertices~\cite{ligra, galois, gap, graphmat, graphlab, graphchi}. Applications may perform pull- or push-based computations. In pull-based computations, a vertex pulls updates from its in-neighbors. In push-based computations, a vertex pushes updates to its out-neighbors. This process may be iterative, and all or only a subset of vertices may participate in a given~iteration.

The \emph{Compressed Sparse Row (CSR)} format is commonly used to represent graphs in a storage-efficient manner. CSR uses a pair of arrays, \emph{Vertex} and \emph{Edge}, to encode the graph. 
CSR encodes in-edges for pull-based computations and out-edges for push-based computations. 
In this discussion, we focus on pull-based computations and note that the observations hold for push-based computation.
For every vertex, the Vertex Array  maintains an index that points to its first in-edge in the Edge Array. 
The Edge Array stores all in-edges, grouped by destination vertex ID. For each in-edge, the Edge Array entry stores the associated source vertex ID. %

The graph applications use an additional \emph{Property} Array(s) to hold partial or final results for every vertex. For example, the \emph{Pagerank} application maintains two ranks for every vertex; one computed from the previous iteration and one being computed in the current iteration. Implementation may use either two separate arrays (each storing one rank per vertex) or may use one array (storing two ranks per vertex).
Fig.~\stitchref{basic-graph}{basic-graph-a}~and~\stitchref{basic-graph}{basic-graph-b} shows a simple graph and its CSR representation for pull-based computations, along with one Property Array.

\subsection{Cache Behavior in Graph Analytics}
\label{general-caching}
\begin{figure}[!t]
    \centering
    {\subfloat{\transparent{1}{\includegraphics[width=1px,height=1px]{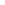}\label{basic-graph-a}}}}
    {\subfloat{\includegraphics[width=0.99\linewidth]{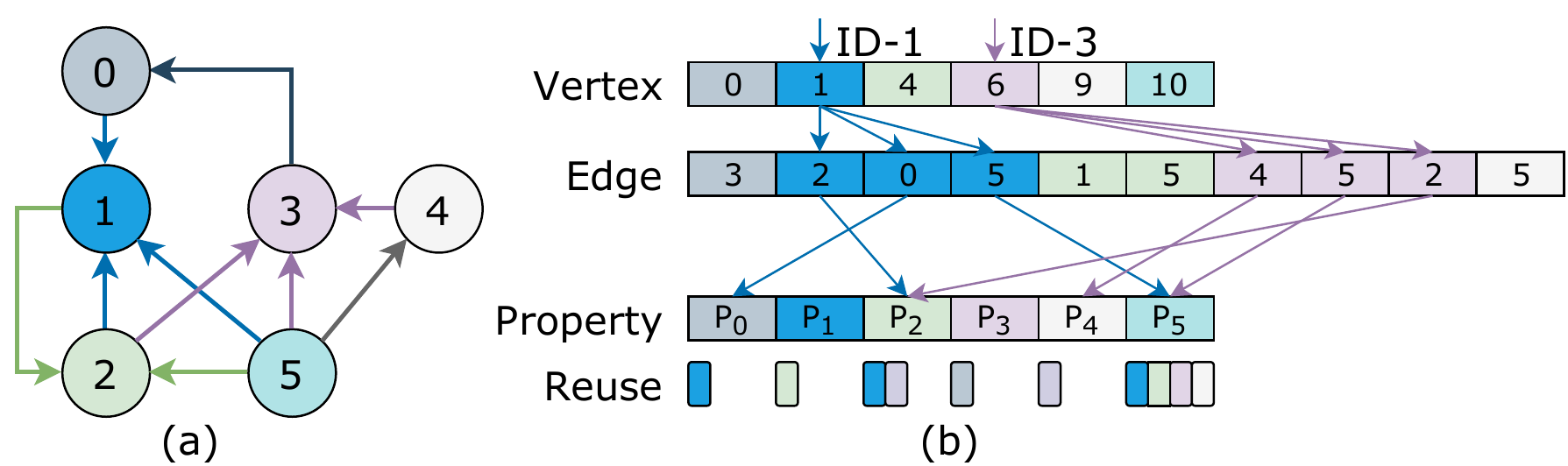}\label{basic-graph-b}}}
    {\subfloat{\transparent{1}{\includegraphics[width=1px,height=1px]{pdfs/transparent.png}\label{basic-graph-c}}}}
    \reducefigcaptionspace
    \caption{(a) An example graph. (b) CSR format encoding in-edges. Elements of the same colors in all arrays, correspond to the same destination vertex. Number of bars (labeled \emph{Reuse}) below each element of the Property Array shows the number of times an element is accessed in one full iteration, where the color of a bar indicates the vertex making an access.
    \label{basic-graph}}
 \end{figure}

 \begin{figure}[!t]
    \ifdefined\RELEASE
        \vspace{.5em}
    \else
    \fi
    \centering
    \includegraphics[width=1\linewidth]{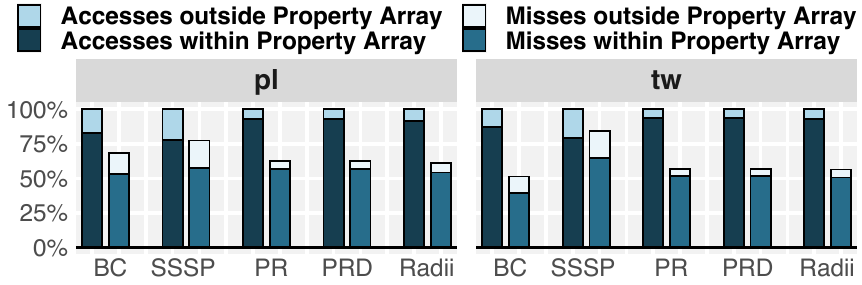}
    \reducefigcaptionspace
    \caption{Classification of LLC accesses and misses (normalized to total accesses) for the \pld{} and \tw{} datasets for the applications from Table~\ref{tab:graph-worklads}.
    }
    \label{fig:breakup} 
    \ifdefined\RELEASE
    \else
        \vspace{-0.5em}
    \fi
\end{figure}

At the most fundamental level, a graph application \mbox{computes} a property for a vertex based on the properties of its neighbours.
To find the neighbouring vertices, an application traverses the portion of the Edge Array corresponding to a given vertex, and then accesses elements of the Property Array corresponding to these neighbouring vertices.
Fig.~\stitchref{basic-graph}{basic-graph-b} highlights the elements accessed during the computations for vertex ID-1 and ID-3. 

As the figure shows, each element in the Vertex and the Edge Array is accessed exactly once during an iteration, exhibiting no temporal locality at LLC. These arrays may exhibit high spatial locality, which is filtered by the L1-D cache, leading to a streaming access pattern in the~LLC.

In contrast, the Property Array does exhibit temporal reuse. However, reuse is not consistent for all elements. %
Specifically, reuse is proportional to the number of out-edges for pull-based algorithms. Thus, the elements corresponding to high out-degree vertices exhibit high reuse. Fig.~\stitchref{basic-graph}{basic-graph-b} shows the reuse for high out-degree (\ie hot) vertices P$_2$ and P$_5$ of the Property Array assuming pull-based computations; other elements do not exhibit reuse. The same observation applies to high in-degree vertices in push-based algorithms.

Finally, Fig.~\ref{fig:breakup} quantifies the LLC behavior of the graph applications listed in Table~\ref{tab:graph-worklads} on the \pld{} and \tw{} datasets as \mbox{representative} examples (Refer to Sec.~\ref{method} for methodology details). The figure differentiates all LLC accesses and misses as falling either within or outside the Property Array. 
Unsurprisingly, the Property Array accounts for 78-94\% of all LLC accesses. However, despite the high reuse, the Property Array is also responsible for a large fraction of LLC misses, the reasons for which are explained next.

\subsection{Challenges in Caching the Property Array}
\label{compute-array}

As discussed in the previous section, elements in the Property Array corresponding to the hot vertices exhibit high reuse. Unfortunately, on-chip caches struggle in capitalizing on the high reuse for the following two~reasons:

\noindentsectiontitle{\bf \smalltextcircled{1}  \em Lack of spatial locality:} the hot vertices are sparsely \mbox{distributed} throughout the memory space of the Property Array. Moreover, the size of each element in the Property Array is much smaller than the size of a cache block. Thus, inevitably, hot vertices share space in a cache block with cold vertices. This leads to cache underutilization as a considerable fraction of a cache block capacity is occupied by the cold~vertices.

\noindentsectiontitle{\bf  \smalltextcircled{2} \em Thrashing in the LLC:} the access pattern to the Property Array is highly irregular, being heavily dependent on both graph structure and application. Between a pair of accesses to a given hot vertex in the Property Array, a number of other, unrelated, cache blocks may be accessed, leading to thrashing. Any block allocated by these unrelated accesses will trigger evictions at the LLC, potentially displacing blocks holding hot~vertices.

\visiblespace
\noindent Overcoming the former problem requires improving cache block utilization by focusing on intra-block reuse, which is effectively addressed by existing software techniques~\cite{dbg,fc,hubcluster}. The latter problem requires retaining high-reuse blocks in the LLC by focusing on reuse across cache blocks; as explained below, this remains an open challenge not addressed by prior works. We note that the two problems are orthogonal in nature; solving one problem does not solve the other one.

We next discuss most relevant state-of-the-art techniques in both software and hardware and their shortcomings in \mbox{addressing} cache thrashing for graph~analytics.

\subsection{Prior Software Schemes}
\label{prior:sw}

Prior works have proposed leveraging application-visible properties, such as vertex connectivity or vertex degree, to improve cache locality. This is accomplished by reordering vertices in memory such that vertices that will be frequently accessed together are kept close by.
These techniques are particularly attractive because they require no modifications to the graph algorithms~\cite{gorder, rabbit, recall, LDG, METIS, SlashBurn, rcm, CHDFS, fc, hubcluster, dbg}. To be effective, these techniques face two constraints during reordering. First, they must keep the reordering cost to a minimum to improve end-to-end application performance. 
Second, they should minimize disruption to the underlying graph structure, specifically for graphs that exhibit community structure.
Prior works have noted that vertex order for many real-world graph datasets closely follow underlying community structure, meaning vertices from the same community are ordered close by in memory, exhibiting good spatio-temporal locality that should be preserved~\cite{fc,hubcluster,dbg}.

While finding an optimal vertex ordering is NP-hard, techniques like \gorder{} attempt to approximate such ordering by comprehensively analyzing the graph structure based on a likely traversal order~\cite{gorder}.
However, recent works show that while such techniques are effective in reducing LLC misses, they 
incur a staggering reordering cost that is often multiple orders of magnitude higher than the application runtime, thus rendering them impractical~\cite{dbg}.

Consequently, the same works argue for lightweight skew-aware techniques that provide application speed-up even after accounting for their reordering cost. To keep the reordering cost low, these techniques reorder vertices solely based on vertex degree, with the goal of improving spatial locality by ensuring that hot vertices share cache blocks only with other hot vertices.
To achieve this effect, skew-aware techniques (\eg HubSort~\cite{fc} and DBG~\cite{dbg}) rely on some variant of degree-based sorting to segregate hot vertices form the cold ones.
While effective in improving spatial locality (as explained in Sec.~\ref{compute-array}), these techniques still suffer from cache thrashing stemming from the fact that the footprint of only hot vertices also exceeds the available cache capacity~\cite{dbg}.

\subsection{Prior Hardware Schemes}
\label{prior:hw}
Prior hardware cache management schemes targeting cache thrashing can be broadly classified into three~categories:

\noindentsectiontitle{\bf \smalltextcircled{1} \em History-agnostic lightweight schemes} use simple \mbox{heuristics} to manage cache~\cite{dip,rrip,tadip,gippr,segmentedLRU,pipp}. RRIP~\cite{rrip} is the state-of-the-art technique in this category that relies on a probabilistic approach to classify a cache block as low- or high-reuse at the time of inserting a new block in the cache. As these techniques do not record the past reuse behavior of cache blocks, they are limited in accurately identifying high-reuse blocks.

\noindentsectiontitle{\bf \smalltextcircled{2} \em History-based predictive schemes} such as the state-of-the-art Hawkeye~\cite{hawkeye} and many others ~\cite{sampler,ship,mdpp,harmony,perceptron,multi,leeway} learn past reuse behavior of cache blocks by employing \mbox{sophisticated} storage-intensive prediction mechanisms. A large body of recent works focus on history-based schemes as they generally provide higher performance than the lightweight schemes for a wide range of applications. However, for graph analytics, we find that graph-dependent irregular access patterns prevent these history-based schemes from correctly learning which cache blocks to preserve. For example, most history-based schemes rely on a PC-based (\emph{Program Counter}) reuse correlation\footnote{All seven schemes~\cite{crc2-1shippp, crc2-2lime, crc2-3multi, crc2-4expected, crc2-5reuse, crc2-6hawkeye, crc2-7red} presented at the Cache Replacement Championship'17~\cite{crc2} rely on PC-based reuse correlation.} to learn which set of PC addresses access high-reuse cache blocks to prioritize these blocks for caching over others. Meanwhile, we observe that the reuse for elements of the Property Array, which are the prime target for LLC caching in graph analytics (Sec~\ref{general-caching}), does not correlate with the PC because the same PC accesses hot and cold vertices~alike.

\noindentsectiontitle{\bf \smalltextcircled{3} \em Pinning-based schemes} such as XMem~\cite{xmem} dedicate partial or full cache capacity by \emph{pinning} high-reuse blocks to cache. Hardware ensures that the pinned blocks cannot be evicted by other cache blocks and thus are protected from cache thrashing. Such an approach is only feasible when the high-reuse working set fits in the available cache capacity. Unfortunately, for large graph datasets, even with high skew, it is unlikely that all hot vertices will fit in the LLC; recall from Table~\ref{tab:hot-vertices-skew} that hot vertices account for up to 26\% of the total vertices. 
Moreover, some of the colder vertices might also exhibit short-term temporal reuse, particularly in graphs with community structure. 

\visiblespace
\noindent These observations call for a new LLC management scheme that employ (1) a reliable mechanism to identify hot vertices amidst irregular access patterns and (2) flexible cache policies that maximize reuse among hot vertices by protecting them in the cache without denying colder vertices the ability to be cache resident if they exhibit reuse.

\section{\sam: Caching in on the Skew}
\label{sec:graphinitydesign}
This work introduces \sam, graph-specialized LLC management for graph analytics processing natural graphs. 
\sam augments existing cache management schemes with simple additions to their insertion and hit-promotion policies that provide preferential treatment to cache blocks containing hot vertices to protect them from thrashing. 
\sam policies are sufficiently flexible to capture reuse of other blocks as~needed.

\samnospace's domain-specialized design is influenced by the following two challenges faced by existing hardware cache management schemes. First, hardware alone cannot enforce spatial locality, which is dictated by vertex placement in the memory space and is under software control. Second, domain-agnostic hardware cache management schemes struggle in pinpointing hot vertices under cache thrashing due to irregular access patterns endemic of graph analytics. 

\begin{figure}[!t]
{
    \centering
    {\subfloat{\transparent{1}{\includegraphics[width=1px,height=1px]{pdfs/transparent.png}\label{fig:graphinity-a}}}}
    {\subfloat{\transparent{1}{\includegraphics[width=1px,height=1px]{pdfs/transparent.png}\label{fig:graphinity-b}}}}
    {\subfloat{\includegraphics[width=0.99\linewidth]{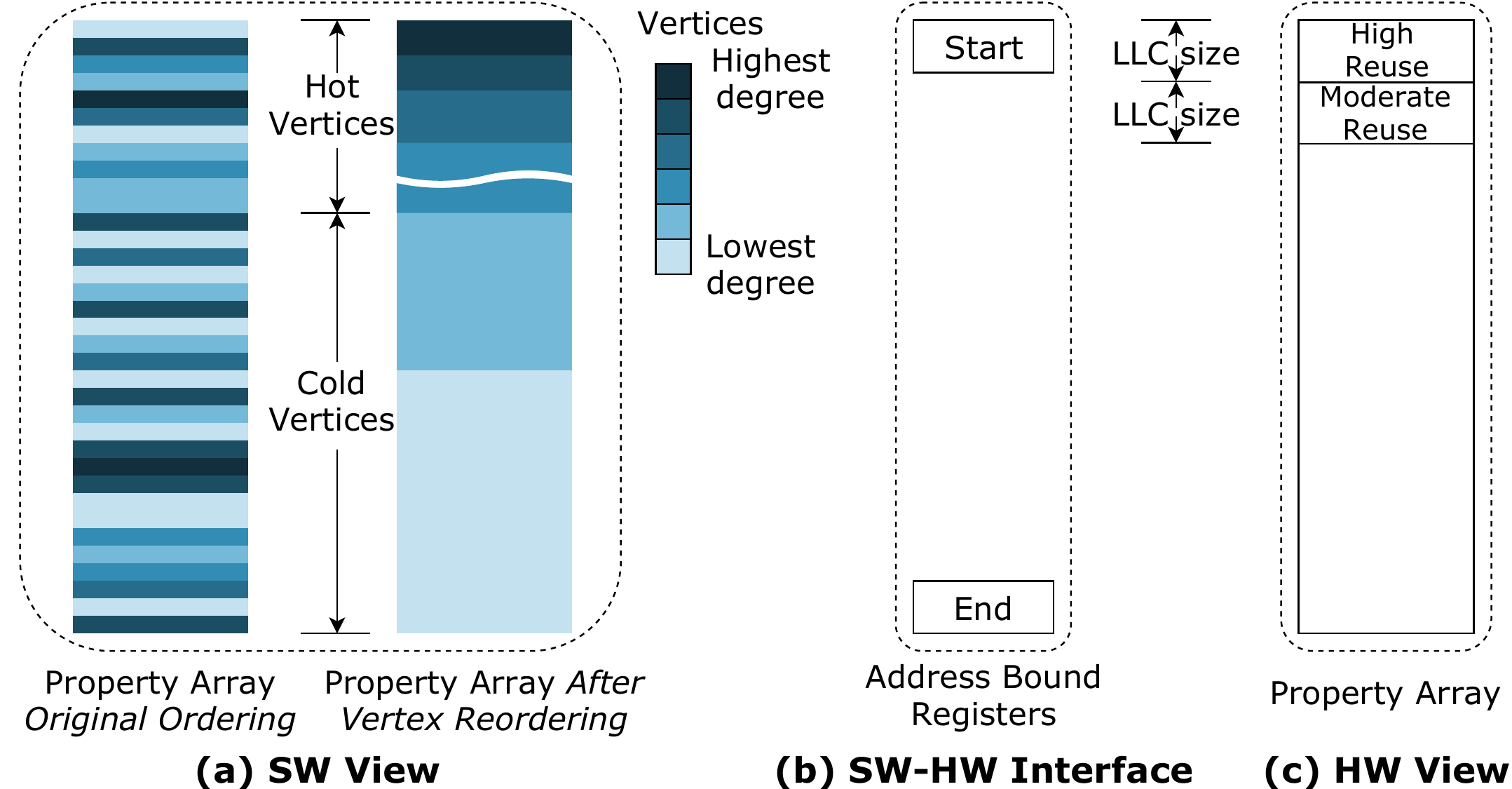}\label{fig:graphinity-c}}}
    \reducefigcaptionspace
    \caption{\sam overview. (a) Software applies vertex reordering, which segregates hot vertices at the beginning of the array. (b) \sam interface exposes an ABR pair per Property Array to be configured with the bounds of the array. (c) \sam identifies regions exhibiting different reuse based on an LLC~size.}
    \label{fig:graphinity}
}
\end{figure}

To overcome both challenges, \sam relies on existing skew-aware reordering software techniques to induce spatial locality by segregating hot vertices in a contiguous memory region~\cite{dbg,fc}. While these techniques offer different trade-offs in terms of reordering cost and their ability to preserve graph structure, they all work by isolating hot vertices from the cold ones. Fig.~\stitchref{fig:graphinity}{fig:graphinity-a} shows a logical view of the placement of hot vertices in the Property Array after reordering by such a technique. 
\sam subsequently leverages the contiguity among hot vertices in the memory space to (1) pinpoint them via a lightweight interface and (2) protect them from thrashing. \sam design consists of three hardware components as follows.

\noindentsectiontitle{\em \smalltextcircled{A} \em Software-hardware interface:} \sam interface is minimal, consisting of a few configurable registers that software populates with the bounds of the Property Array during the initialization of an application (see Fig.~\stitchref{fig:graphinity}{fig:graphinity-b}). Once populated, \sam does not rely on any further intervention from software. 

\noindentsectiontitle{\em \smalltextcircled{B} \em Classification logic:} \sam logically partitions the Property Array into different regions  based on expected reuse.
(See Fig.~\stitchref{fig:graphinity}{fig:graphinity-c}). \sam implements simple comparison-based logic, which, at runtime, classifies whether a cache request belongs to one of these regions.

\noindentsectiontitle{\em \smalltextcircled{C} \em Specialized cache policies:} \sam specializes cache policies for each region to ensure hot vertices are protected from thrashing while retaining flexibility in caching other blocks. The classification logic informs the choice of which policy to apply to a given cache block.

\begin{figure}[!t]
{
    \vspace{-0.75em}
    \centering
    \ifdefined\RELEASE
    \includegraphics[width=1\linewidth]{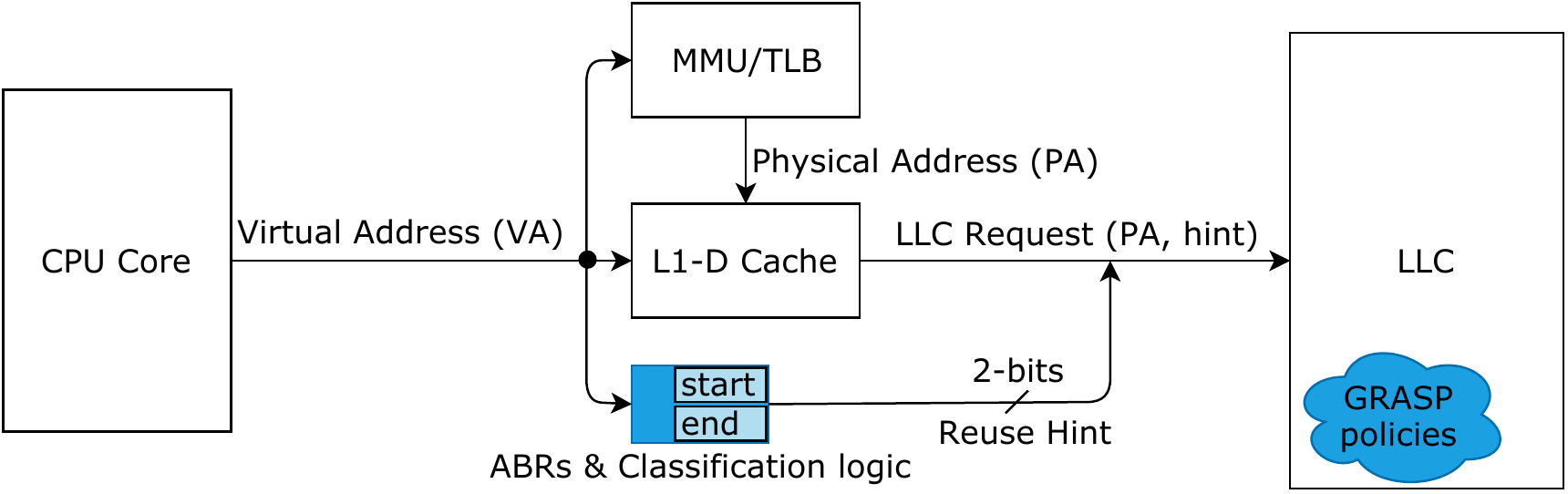}%
    \else
    \includegraphics[width=0.8\linewidth]{pdfs/SAMBlockDiagram.pdf}%
    \fi
    \vspace{-0.5\baselineskip}
	\caption{Block diagram of \sam and other hardware components with which it interacts. \sam components are shown in color. For brevity, the figure shows only one CPU core.}
	\label{fig:sam-block-diagram}
	\ifdefined\RELEASE
    \vspace{-0.5em}
    \else
    \vspace{-1em}
    \fi
}
\end{figure}

\noindentsectiontitle{}Fig.~\ref{fig:sam-block-diagram} shows how \sam interacts with other hardware components in the system. In the following sections, we describe each of \samnospace's components in detail.


\subsection{Software-Hardware Interface\label{interface}}

\samnospace's interface consists of one pair of \emph{Address Bound Registers (ABR)} per Property Array; recall from Sec.~\ref{graph-processing-basics} that an application may maintain more than one Property Array, each of which requires a dedicated ABR pair. ABRs are part of an application context and are exposed to the software.
At application start-up, the graph framework populates each ABR pair with the start and end virtual address of the entire Property Array (Fig.~\stitchref{fig:graphinity}{fig:graphinity-b}). 
Setting these registers activates the custom cache management for graph analytics. 
When the ABRs are not set by the software (\ie the default case for other applications), specialized cache management is essentially disabled.

The use of virtual addresses keeps the GRASP interface independent of the existing TLB design, allowing GRASP to perform address classification (described next) in parallel with the usual virtual-to-physical address translation carried out by TLB (see Fig.~\ref{fig:sam-block-diagram}).
Prior works have used similar approaches to pass data-structure bounds to aid microarchitecture mechanisms~\cite{prefetch-data-structure,barren,whirlpool,xmem}.

\subsection{Classification Logic}

This component of \sam is responsible for reliably identifying cache blocks containing hot vertices in hardware by leveraging the bounds of the Property Array(s) available in the ABRs as explained in the following sections:

\noindentsectiontitle{\bf \em Identifying Hot Vertices: } In theory, all hot vertices should be cached. In practice, it is unlikely that all hot vertices will fit in the LLC for large datasets.
In such a case, providing preferential treatment to all hot vertices is {\em not} beneficial as they can thrash each other in the LLC.
To avoid this problem, \sam prioritizes cache blocks containing only a subset of hot vertices, comprised of only the hottest vertices based on available LLC capacity. Conveniently, the hottest vertices are located at the beginning of the  Property Array in a contiguous region thanks to the application of skew-aware reordering as seen in Fig.~\stitchref{fig:graphinity}{fig:graphinity-a}.

\noindentsectiontitle{\bf \em Pinpointing the High Reuse Region: }
\sam labels two LLC-sized sub-regions within the Property Array: 
The LLC-sized memory region at the start of the Property Array is labeled as \emph{High Reuse Region}; another LLC-sized memory region starting immediately after the High Reuse Region is labeled as the \emph{Moderate Reuse Region} (Fig.~\stitchref{fig:graphinity}{fig:graphinity-c}). Finally, if an application specifies more than one Property Array, \sam divides LLC-size by the number of Property Arrays before labeling the~regions.  

\noindentsectiontitle{\bf \em Classifying LLC Accesses: }
At runtime, \sam classifies a memory address making an LLC access as \emph{High-Reuse} if the address belongs to the {High Reuse Region} of any Property Array; \sam determines this by comparing the address with the bounds of the High Reuse Region of each Property Array. Similarly, an address is classified as \emph{Moderate-Reuse} if the address belongs to the {Moderate Reuse Region}.
All other LLC accesses are classified as \emph{Low-Reuse}. 
For non-graph applications, the ABRs are not initialized and all accesses are classified as \emph{Default}, effectively disabling domain-specialized cache management. \sam encodes the classification result (High-Reuse, Moderate-Reuse, Low-Reuse or Default) as a 2-bit \emph{Reuse Hint}, and forwards it to the LLC along with each cache request, as shown in Fig.~\ref{fig:sam-block-diagram}, to guide specialized insertion and hit-promotion policies as described next.

\subsection{Specialized Cache Policies\label{sam}}%
This component of \sam implements specialized cache policies that protect the cache blocks associated with High-Reuse LLC accesses against thrashing. One naive way of doing so is to pin the High-Reuse cache blocks in the LLC. However, pinning would sacrifice any opportunity in exploiting temporal reuse that may be exposed by other cache blocks (\eg Moderate-Reuse cache blocks).

To overcome this challenge, \sam adopts a flexible approach by augmenting an existing cache replacement policy with a specialized insertion policy for LLC misses and a hit-promotion policy for LLC hits. \samnospace's specialized policies provide preferential treatment to High-Reuse blocks while maintaining flexibility in exploiting temporal reuse in other cache blocks, as discussed next.

\noindentsectiontitle{\bf \em Insertion Policy:} Accesses tagged as {High-Reuse}, comprising the set of the hottest vertices belonging to the High Reuse Region, are inserted in the cache at the MRU position to protect them from thrashing. 
Accesses tagged as {Moderate-Reuse}, likely exhibiting lower reuse when compared to the High-Reuse region, are inserted near the LRU position. Such insertion policy allows Moderate-Reuse cache blocks an opportunity to experience a hit without causing thrashing. Finally, accesses tagged as {Low-Reuse}, comprising the rest of the graph dataset, including the long tail of the Property Array containing cold vertices, are inserted at the LRU position, thus making them immediate candidates for replacement while still providing them with an opportunity to experience a hit and be promoted using the specialized policy described~next. 

\noindentsectiontitle{\bf \em Hit-Promotion Policy:} Cache blocks associated with {High-Reuse} LLC accesses are immediately promoted to the MRU position on a hit to protect them from thrashing. LLC hits to blocks classified as {Moderate-Reuse} or {Low-Reuse} make for an interesting case. 
On the one hand, the likelihood of these blocks having further reuse is quite limited, which means they should not be promoted directly to the MRU position. On the other hand, by experiencing at least one hit, these blocks have demonstrated temporal locality, which cannot be completely ignored. \sam takes a middle ground for such blocks by gradually promoting them towards MRU position on every hit.

\noindentsectiontitle{\bf \em Eviction Policy:} \samnospace's eviction policy does not utilize hint to \mbox{differentiate} blocks at replacement time; hence, it is unmodified from the baseline scheme. This is a key factor that keeps the cache management flexible for \samnospace. By not prioritizing candidates for eviction, \sam ensures that blocks classified as {High-Reuse} but not referenced for a long time can yield cache space to other blocks that do exhibit reuse. Because the unchanged eviction policy does not need to differentiate between blocks with High-Reuse and other hints, cache blocks do \emph{not} need to explicitly store the Reuse Hint as additional LLC metadata.

\visiblespace

\begin{table}[!t]
    \caption{\footnotesize{Policy columns show how \sam updates per-block 3-bit RRPV counter of RRIP (base scheme) for a given Reuse Hint. Higher RRPV value indicates higher eviction priority.}}
    \label{sam-rrip}
    \reducetablecaptionspace
    \centering
    \footnotesize
    \begin{tabular}{|l||l|l|}
        \hline
        {\bf Reuse Hint} & \multicolumn{1}{l|}{\bf Insertion Policy} & \multicolumn{1}{l|}{\bf Hit Policy} \\ \hline
        \hline
        High-Reuse  & RRPV = 0  & RRPV = 0    \\ \hline
        Moderate-Reuse & RRPV = 6 & if RRPV $>$ 0:    \\ \cline{1-2}
        Low-Reuse   & RRPV = 7 & \multicolumn{1}{c|}{RRPV - -}    \\ \hline
        Default   & RRPV = 6 or 7 & RRPV = 0    \\ \hline
    \end{tabular}
\end{table}

Table~\ref{sam-rrip} shows the specialized cache policies for all Reuse Hints under \samnospace. While the table, and our evaluation, assumes RRIP~\cite{rrip} as the base replacement scheme, we note that \sam is not fundamentally dependent on RRIP and can be implemented over many other schemes including, but not limited to, LRU, Pseudo-LRU and DIP~\cite{dip}. %

\subsection{Benefits of \sam over Prior Schemes}
The state-of-the-art history-based schemes
~\cite{sampler,mdpp,ship,perceptron,hawkeye,harmony,leeway,multi} 
require intrusive modifications to the cache structure in form of embedded metadata in cache blocks and/or dedicated predictor tables. These schemes also require propagating a PC signature through the core pipeline all the way to the LLC, which so far has hindered their commercial~adoption. 

In comparison, \sam is implemented within the same hardware structure required by the base lightweight scheme (e.g., RRIP). 
\sam propagates only a \mbox{2-bit} {Reuse Hint} to the LLC on each cache access to guide cache policy decisions. %
By relying on lightweight software support, \sam reliably pinpoints hot vertices in hardware without requiring costly prediction tables and/or additional per-cache-block metadata. 

When compared to pinning-based schemes, \sam \mbox{policies} protect hot vertices from thrashing while remaining flexible to capture reuse of other blocks as needed.
Combining robust cache policies with minimal hardware modifications makes \sam feasible for commercial adoption while also providing higher LLC~efficiency.

\section{Methodology}
\label{method}

\begin{table}[!t]
    \caption{\footnotesize{Graph-analytic applications.}}
    \label{tab:graph-worklads}
    \reducetablecaptionspace
    \centering
    \footnotesize
    \begin{tabularx}{1\linewidth}
        {|>{\centering\arraybackslash\hsize=0.20\hsize}X|| 
          >{\raggedright\arraybackslash\hsize=0.80\hsize}X|
        }
        \hline
        {\bf Application}    & \multicolumn{1}{c|}{\bf Brief description} \\ \hline
        \hline
        {Betweenness Centrality ~~~~~(BC)} & finds the most central vertices in a graph by using a BFS kernel to count the number of shortest paths passing through each vertex from a given root vertex. \\ \hline
        {Single~Source Shortest Path (SSSP)} & computes shortest distance for vertices in a weighted graph from a given root vertex using the Bellman-Ford algorithm. \\ \hline
        {Pagerank ~~~~~(PR)}  & is an iterative algorithm that calculates ranks of vertices based on the number and quality of incoming edges to them~\cite{pagerank}. \\ \hline
        {PageRank-Delta (PRD)} &is a faster variant of PageRank in which vertices are active in an iteration only if they have accumulated enough change in their PageRank score. \\ \hline %
        {Radii Estimation (Radii)} & estimates the radius of each vertex by performing multiple parallel BFS's from a small sample of vertices~\cite{radii}. \\
        \hline
    \end{tabularx}
\end{table}

\begin{table}[!t]
    \caption{\footnotesize{Effect of our optimization on the original Ligra implementation for different applications. PR applies pull-based computations whereas SSSP applies push-based computations throughout the execution; the rest of the applications switch between pull or push based on a number of active vertices in a given iteration.}}
    \label{tab:opt}
    \reducetablecaptionspace
    \centering
    \footnotesize
    \begin{tabular}{|c||c|c|}
        \hline
        \multicolumn{1}{|c||}{\bf Application} &  \multicolumn{1}{c|}{\bf Merging opportunity?} & \multicolumn{1}{c|}{\bf Speed-up} \\
        \hline \hline
        \bc{} & No & - \\ \hline
        \sssp{} & Yes & 3-8\%\\ \hline
        \pr{} & Yes & 40-52\%\\ \hline
        \prd{} & Yes & 14-49\%\\ \hline
        \radii{} & No & - \\ \hline
    \end{tabular}
\end{table}

\subsection{Graph Processing Framework}
\label{sec:Applications}
For the evaluation, we use \emph{Ligra}~\cite{ligra}, a widely used graph processing framework that supports both pull- and push-based computations, including switching from pull to push (and vice versa) at the start of every iteration. We combine the five diverse applications listed in Table~\ref{tab:graph-worklads} with the five high-skew graph datasets listed in Table~\ref{tab:datasets}, resulting in 25 benchmarks. To test the robustness of \sam to adversarial workloads, we use two additional datasets with low-/no-skew.

We obtained the source code for the graph applications from Ligra~\cite{ligra} 
and applied a simple data-structure \mbox{optimization} to improve locality in the baseline implementation as follows.
As explained in Sec.~\ref{general-caching}, graph applications exhibit irregular accesses for the Property Array, with applications potentially maintaining more than one such array. When multiple Property arrays are used, elements corresponding to a given vertex may need to be sourced from all of the arrays. We
merge these arrays to induce spatial locality, which reduces number of misses, and in turn, improves performance on all datasets for \pr{}, \prd{} and \sssp{} (see Table~\ref{tab:opt}). 
We use the optimized implementation of these three applications as a stronger baseline for our evaluation. The optimized applications are available at \url{https://github.com/faldupriyank/grasp}. We do note that GRASP does not mandate merging arrays as GRASP design can accommodate multiple arrays. Nevertheless, merging does reduce the number of arrays needed to be tracked.

For \prd, two versions of the algorithm are provided with Ligra: push-based and pull-push. In the baseline \mbox{implementation}, the push-based version is faster. However, after merging the Property Arrays, the pull-push variant performs better, and is what we use for the evaluation. 

\setlength\tabcolsep{5pt}
\subsection{Methodology for Software Evaluation \label{sec:sw-eval-method}}
\begin{table}[!t]
    \caption{\footnotesize{Properties of the graph datasets. Top five datasets are used in the main evaluation whereas the bottom two datasets are used as adversarial datasets.}}
    \label{tab:datasets}
    \reducetablecaptionspace
    \centering
    \footnotesize
    \begin{tabular}{|r||r|r|c|}
    \hline
    { \bf Dataset } & 
    \multicolumn{1}{c|}{\bf Vertex Count} & 
    \multicolumn{1}{c|}{\bf Edge Count}  & 
    \multicolumn{1}{c|}{\bf Avg. Degree} \\
    
    \hline
    \hline
    { LiveJournal ({\lj})~\cite{snapnets}}
    & \num{5}{M}
    & \num{68}{M}
    & 14 \\ \hline 
    
    { PLD ({\pld})~\cite{pld}}            
    & \num{43}{M}             
    & \num{623}{M}       
    & 15 \\ \hline
    
    {Twitter ({\tw})~\cite{twitter}}      
    & \num{62}{M}             
    & \num{1468}{M}       
    & {24} \\ \hline

    { Kron ({\kr})~\cite{gap}}            
    & \num{67}{M}             
    & \num{1323}{M}       
    & {20} \\ \hline
    
    { SD1-ARC ({\sd})~\cite{pld}}            
    & \num{95}{M}             
    & \num{1937}{M}       
    & {20} \\ \hline \hline

    {Friendster ({\fr}) \cite{konect-friendster}}   & 
    \num{64}{M} & 
    \num{2147}{M} & 
    {33} \\ \hline
    
    {Uniform ({\uni}) \cite{rmat}}   & 
    \num{50}{M} & 
    \num{1000}{M} & 
    {20} \\
    
    \hline
    \end{tabular}
\end{table}

\noindentsectiontitle{\bf \em Server Configuration:} Native experiments (Sec.~\ref{eval-reordering-techniques}) are performed on a dual-socket server with two Broadwell based {\em Intel Xeon CPU E5-2630}~\cite{xeon}, each with 10 cores clocked at 2.2GHz and a 25MB shared LLC. Hyper-threading is enabled, exposing 40 hardware execution contexts across both CPUs. The machine has 128GB of DRAM provided by eight DIMMs clocked at 2133MHz. All experiments were run using 40 threads, and we pinned the software threads to avoid performance variations due to OS scheduling. To further reduce sources of performance variation, we also disable the \emph{turbo boost} DVFS features. Finally, we enabled \emph{Transparent Huge Pages} to reduce TLB misses.

Evaluation of software reordering techniques are carried out on the server mentioned above. We report the performance speed-up over the entire application runtime (including reordering cost) but exclude the graph loading time from the disk. For iterative applications, \pr{} and \prd{}, we run them until convergence and consider the runtime over all iterations. For root-dependent traversal applications, \sssp{} and \bc{}, we run them from eight different root vertices for each input dataset and consider the runtime over all eight traversals.
Finally, we run six executions for each application-dataset pair and report the geometric mean of the five trials, excluding the timing of the first trial to allow the caches to warm up. We note that runtime is relatively stable across executions; for each reported datapoint, coefficient of variation is below~2\%.

\noindentsectiontitle{\bf \em Reordering Techniques:} 
We evaluate the following reordering techniques and use the source code 
for skew-aware techniques
from \url{https://github.com/faldupriyank/dbg}
and 
for Gorder 
from \url{https://github.com/datourat/Gorder}.

\noindentsubsectiontitle{\bf Sort} reorders vertices in the memory space by sorting them in the descending order of their degree.

\noindentsubsectiontitle{\bf HubSort}~\cite{fc}  segregates hot vertices in a contiguous region by assigning them a continuous range of vertex IDs in their descending order of degree. In doing so, Hub Sorting essentially \emph{sorts all hot vertices}, while largely preserving structure for the cold vertices.

\noindentsubsectiontitle{\bf DBG}~\cite{dbg}, unlike Sort and HubSort, does not rely on sorting to segregate hot vertices. Instead, DBG coarsely partitions all vertices into a small number of groups based on their degree. Similar to Sort and HubSort, DBG is effective at improving spatial locality; however, unlike the other two techniques, DBG is able to largely preserve the existing graph structure.

\noindentsubsectiontitle{\bf \gorder}~\cite{gorder} is evaluated as a representative of complex \mbox{techniques}. As Gorder is only available in a single-thread implementation, while reporting the net runtime of Gorder for a given dataset, we optimistically divide the reordering time by 40 (maximum number of threads supported on the server) to provide a fair comparison with skew-aware techniques whose reordering implementation is fully parallelized.

\subsection{Methodology for Hardware Evaluation\label{sec:hw-eval-method}}

\noindentsectiontitle{\bf \em Simulation Infrastructure:} We use the {\em Sniper}~\cite{sniper} simulator modeling 8 OoO cores. Table~\ref{sim-params} lists the parameters of the simulated system. The applications are evaluated in a multi-threaded mode with 8-threads.

We find that the graph applications spend significant fraction (86\% on average in our evaluations) of time in push-based iterations for \sssp{} or pull-based iterations for all other evaluated applications. Thus, we simulate the \emph{Region of Interest (ROI)} covering only push- or pull-based iterations (whichever one dominates) for the respective applications.
Because simulating all iterations of a graph-analytic application in a detailed microarchitectural simulator is prohibitive, time-wise, we instead simulate one iteration that has the highest number of active vertices. To validate the soundness of our methodology, we also simulated one more randomly chosen iteration for each application-dataset pair with at least 20\% of vertices active and observed trends similar to the ones reported in the paper.

\noindentsectiontitle{\bf \em Hardware Cache Management Schemes:} 
\begin{table}[!t]
    \caption{\footnotesize{Parameters of the simulated system for evaluation of the hardware schemes.}}
    \label{sim-params}
    \reducetablecaptionspace
    \centering
    \footnotesize
    \begin{tabular} {|r||l|}%
        \hline
        Core        &   OoO @ 2.66GHz, 4-wide front-end\\ \hline%
        \multirow{2}{*}{L1-I/D Cache}     &   4/8-ways 32KB, 4 cycles access latency   \\
        & stride-based prefetchers with 16 streams \\ \hline
        L2 Cache    &   Unified, 8-ways 256KB, 6 cycles access latency  \\ \hline            
        \multirow{2}{*}{L3 Cache}    &   16-ways 16MB NUCA (2MB slice per core),\\
                    &   10 cycles bank access latency \\ \hline            
        NOC         &   Ring network with 2 cycles per hop \\ \hline            
        Memory      &   50ns latency, 2 on-chip memory controllers \\ \hline
    \end{tabular}
\end{table}
We evaluate \sam and compare it with the state-of-the-art thrash-resistant cache management schemes described below.

\noindentsubsectiontitle{\bf RRIP}~\cite{rrip} is the state-of-the-art lightweight scheme that does not depend on history-based learning. RRIP is the most appropriate comparison point given that \sam builds upon RRIP as the base scheme (Sec.~\ref{sam}). We implement RRIP (specifically, {\em DRRIP}) based on the source code from the cache replacement championship (CRC1)~\cite{championship} for RRIP, and use a 3-bit counter per cache block. We use RRIP as a high-performance baseline and report speed-up for all hardware schemes over the RRIP baseline (except for the studies in Sec~\ref{sec:oracle} that use LRU baseline).

\noindentsubsectiontitle{\bf Signature-based Hit Predictor (SHiP)}~\cite{ship} is the state-of-the-art insertion policy which builds on RRIP~\cite{rrip}. Due to the shortcomings of PC-based reuse correlation for graph applications as explained in Sec.~\ref{prior:hw}, we evaluate a SHiP-MEM variant that correlates a block's reuse to the block's memory region. We evaluate 16KB memory regions as in the original proposal. The predictor table is provisioned with an {\em unlimited number of entries} to assess the maximum potential of the scheme.
Each table entry consists of a 3-bit saturating counter that tracks the re-reference behavior of cache blocks of the memory region associated with that entry.

\noindentsubsectiontitle{\bf Hawkeye}~\cite{hawkeye} is the state-of-the-art cache management scheme and winner of the cache replacement championship (CRC2)~\cite{crc2}. Hawkeye trains its predictor table by applying Belady's MIN algorithm on past LLC accesses to infer block's cache friendliness. We use the source code from the CRC2 for Hawkeye that improves upon the prefetcher-agnostic design of Hawkeye from~\cite{hawkeye}. We appropriately scale the number of sampling sets and predictor table entries for a 16MB cache. 

\noindentsubsectiontitle{\bf Leeway}~\cite{leeway}
is a history-based cache management scheme that applies dead block predictions based on a metric called Live Distance, which conservatively captures the reuse interval of a cache block. 
We use the most recent version of the source code for Leeway from
\url{https://github.com/faldupriyank/leeway}.
We appropriately scale the number of sampling sets and predictor table entries for a 16MB cache.

\begin{figure*}[!t]
    \centering
    \includegraphics[width=1\linewidth]{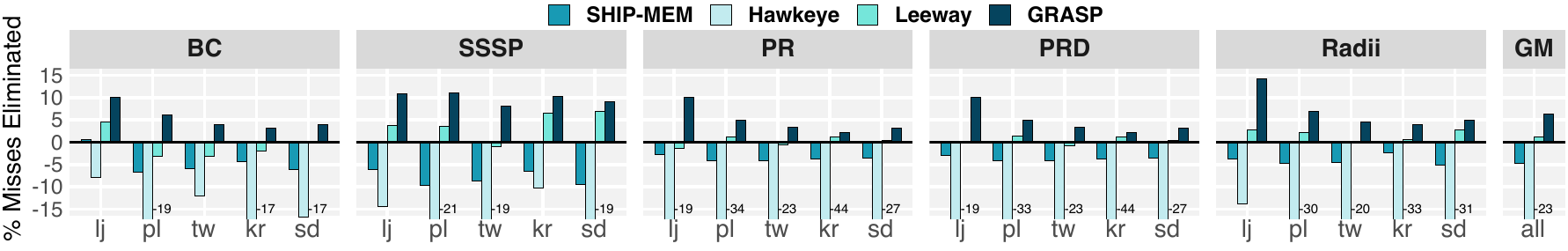}
    \reducefigcaptionspace
    \caption{LLC miss reduction for \sam and state-of-the-art history-based cache management schemes over the RRIP baseline.}
    \label{fig:sam-main1-miss} 
\end{figure*}

\begin{figure*}[!t]
    \centering
    \includegraphics[width=1\linewidth]{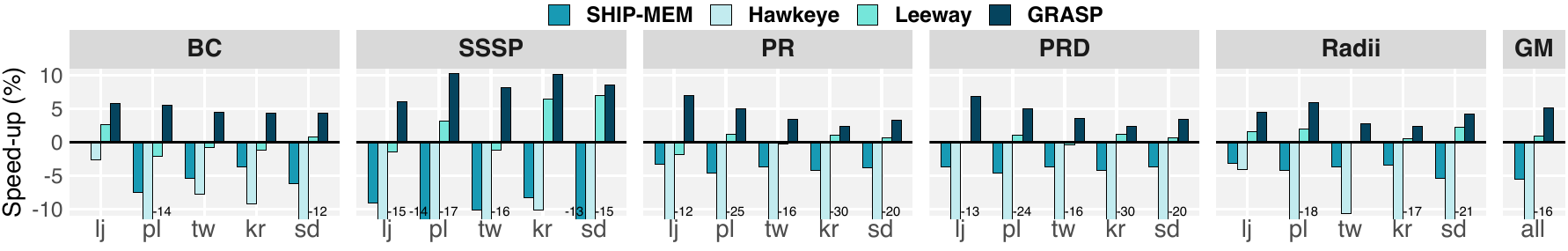}
    \reducefigcaptionspace
    \caption{Speed-up for \sam and state-of-the-art history-based cache management schemes over the RRIP baseline.}
    \label{fig:sam-main1-perf}
\end{figure*}

\noindentsubsectiontitle{\bf XMem}~\cite{xmem} is a pinning-based scheme, originally proposed for \mbox{algorithms} that benefit from {\em cache tiling}. A cache block, once pinned, cannot be evicted until explicitly unpinned by the software, usually done when the processing of a tile is complete. 
In the original proposal, XMem reserves 75\% of LLC capacity to pin tile data whereas the remaining capacity is managed by the base replacement scheme for the rest of the data. In this work, we explore four configurations of XMem, labeled PIN-X, where X refers to the percentage (25\%, 50\%, 75\% or 100\%) of LLC capacity reserved for pinning.
We adopt XMem design for graph analytics and identify the cache blocks from the high reuse region that benefit from pinning using the \sam interface.
Finally, XMem requires an additional 1-bit for every cache block to identify whether a cache block is pinned, along with an additional mechanism to track how much of the capacity is used by the pinned cache blocks at any given time. 

\noindentsubsectiontitle{\bf \sam} is the proposed domain-specialized cache \mbox{management} scheme for graph analytics. We instrument the applications to communicate the address bounds of the Property Arrays to the simulated \DBCM hardware. For each evaluated application, we needed to instrument at most two arrays.
Finally, \sam uses RRIP as the base cache scheme with a 3-bit saturating counter and does {\em not} add any further storage to per-block metadata.

\section{Evaluation\label{sec:eval}}

We first evaluate hardware cache management schemes on top of a skew-aware software reordering technique (Sec.~\ref{sec:eval-predictive} \&~\ref{sec:eval-pin}). Due to long simulation time, evaluating all hardware schemes on top of all four software reordering techniques would be prohibitive. Thus, without loss of generality, we evaluate hardware schemes on top of DBG, which consistently outperforms other reordering techniques (Sec.~\ref{eval-reordering-techniques}). In Sec.~\ref{sam-compatibility}, we evaluate \sam with other reordering techniques to show GRASP's generality.

\subsection{History-based Predictive Schemes\label{sec:eval-predictive}}

In this section, we compare GRASP with the state-of-the-art hardware schemes, SHiP-MEM~\cite{ship}, Hawkeye~\cite{hawkeye} and Leeway~\cite{leeway}.
Prior cache management proposals typically used LRU as the baseline, which is known to be inefficient against thrashing access patterns.
Thus, we use RRIP, the state-of-the-art history-agnostic caching scheme as a strong baseline.
Finally, we use DBG as the software baseline; thus, all speed-ups reported in this section are \emph{over and above~DBG}.

\noindentsectiontitle{\bf \em Miss reduction:} Fig.~\ref{fig:sam-main1-miss} shows the miss reduction over the RRIP baseline. \sam consistently reduces misses on all datapoints, eliminating 6.4\% of LLC misses on average and up to 14.2\% in the best case (on \lj{} dataset for the Radii application). 
The domain-specialized design allows \sam to accurately identify the high-reuse working set (\ie hot vertices), which \sam is able to retain in the cache through its specialized policies, effectively exploiting the temporal~reuse.%

Among prior techniques, Leeway is the only technique that reduces misses, albeit marginal, with an average miss reduction of 1.1\% over the RRIP baseline. The other two techniques are not effective for graph applications, with SHiP-MEM and Hawkeye \emph{increasing} misses across datapoints, with an average miss reduction of -4.8\% and -22.7\%, respectively, over the baseline. This is a new result as prior works show that Hawkeye and SHiP-MEM outperform RRIP on a wide range of applications~\cite{ship,hawkeye}.  
The result indicates that the learning mechanisms of the state-of-the-art domain-agnostic schemes are deficient in retaining the high-reuse working set for graph applications, which ends up hurting application performance as discussed next.

\noindentsectiontitle{\bf \em Application speed-up:}
Fig.~\ref{fig:sam-main1-perf} shows the speed-up for hardware schemes over the RRIP baseline. Overall, performance correlates well with the change in LLC misses; 
\sam consistently provides a speed-up across datapoints with an average speed-up of 5.2\% and up to 10.2\% in the best case (on \pld{} dataset for the SSSP application) over the baseline. When compared to the same baseline, SHiP-MEM and Hawkeye consistently cause slowdown with an average speed-up of -5.5\% and -16.2\%, respectively whereas Leeway yields a marginal speed-up of 0.9\%.
Finally, when compared to prior works directly, \sam yields 4.2\%, 5.2\%, 11.2\% and 25.5\% average speed-up over Leeway, RRIP, SHiP-MEM and Hawkeye, respectively, while not causing slowdown on any datapoints. %

We also evaluated prior schemes without applying any vertex reordering. Leeway, SHiP-MEM and Hawkeye yield an average speed-up of -0.8\%, -5.7\% and \mbox{-14.8\%}, respectively, over RRIP on the datasets with no reordering. However, detailed study is omitted due to space constraints.

\noindentsectiontitle{\bf \em Dissecting Performance of SHiP-MEM:} SHiP-MEM is a history-based scheme that predicts reuse of a cache block based on the fine-grained memory region, to which it belongs. Thus, SHiP-MEM relies on a homogeneous cache behavior for all blocks belonging to the same memory region. In theory, DBG (or another skew-aware technique) should allow SHiP-MEM to identify memory regions containing hottest of vertices (corresponding to High Reuse Region from Fig.~\stitchref{fig:graphinity}{fig:graphinity-c}). In practice, irregular access patterns to these regions and thrashing by cache blocks from other regions impede learning.
Thus, despite leveraging software and utilizing a sophisticated storage-intensive prediction mechanism in hardware, SHiP-MEM underperforms domain-specialized \samnospace.

\noindentsectiontitle{\bf \em Dissecting Performance of Hawkeye:} Hawkeye is the state-of-the-art history-based scheme that uses PC-based reuse correlation to predict whether a cache block has a cache-friendly or cache-averse behavior based on past LLC accesses.
Thus, Hawkeye fundamentally relies on homogeneous cache behavior for all blocks accessed by the same PC address.
When Hawkeye is employed for graph analytics, Hawkeye struggles to learn the behavior of cache blocks in the Property Array as hot vertices exhibit cache-friendly behavior while cold vertices exhibit cache-averse behavior, yet all vertices are accessed by the same PC address.
To make matters worse, if a block incurs a hit and Hawkeye predicts the PC making the access as cache-averse, the cache block is prioritized for eviction instead of promoting the block to MRU as is done in the baseline. 
Thus, Hawkeye performs even worse than the baseline for all combinations of graph applications and datasets.
While not evaluated, other PC-based schemes (e.g.,~\cite{sampler,ship}) that rely on a PC-based correlation would also struggle on graph applications for the same reason.

\noindentsectiontitle{\bf \em Dissecting Performance of Leeway:}
Leeway, like Hawkeye, also relies on a PC-based reuse correlation, and thus is not expected to provide significant speed-ups for graph-analytics. However, Leeway successfully avoids the slowdown on 10 of the 25 datapoints and significantly limits the slowdown on the rest of the datapoints (max slowdown of 2.1\% vs 13.6\% for SHiP-MEM and 30.2\% for Hawkeye). The reasons why Leeway perfroms better than prior PC-based schemes can be attributed to (1) the conservative nature of the Live Distance metric, which Leeway uses to determine if a cache block is dead, and (2) adaptive reuse-aware policies that control the rate of predictions based on the observed access patterns. Because of these two factors, performance of Leeway remains close the the base replacement scheme in the presence of variability in cache blocks reuse behavior. 

\begin{figure}[!t]
    \centering
    \includegraphics[width=1\linewidth]{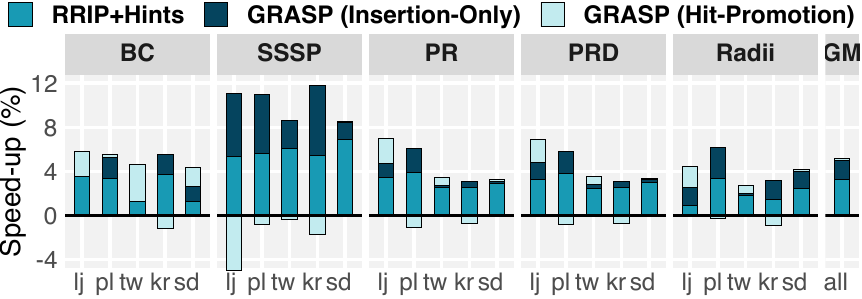}
    \reducefigcaptionspace
    \caption{Impact of GRASP features on performance.}
    \label{fig:grasp-policies}
\end{figure}

\noindentsectiontitle{\bf \em Dissecting Performance of GRASP:}
Performance of GRASP over its base scheme, RRIP, can be attributed to three features: software hints, insertion policy and hit-promotion policy. 
Fig.~\ref{fig:grasp-policies} shows the performance impact due to each of these features. 
RRIP inserts every new cache block at one of the two positions (as specified in the Default Reuse Hint of Table~\ref{sam-rrip}); a cache block is inserted at the LRU position with a high probability or near the LRU position with a low probability.
RRIP+Hints is identical to RRIP except for how a new cache block is assigned these positions. RRIP+Hints uses software hints (similar to GRASP) to guide the insertion. A cache block with High-Reuse hint is inserted near the LRU position and all other blocks are inserted at the LRU position.
GRASP (Insertion-Only) refers to the scheme that applies insertion policy of GRASP as specified in Table~\ref{sam-rrip} but the hit-promotion policy is unchanged from RRIP. 
Finally, GRASP (Hit-Promotion) refers to the scheme that applies hit-promotion policy of GRASP along with its insertion policy, which is essentially the full GRASP design. Note that each successive scheme adds a new feature on top of the features incorporated by the previous ones. For example, GRASP (Insertion-Only) features a new insertion policy in addition to the software hints.

As the figure shows, RRIP+Hints yields an average speed-up of 3.3\% over probabilistic RRIP, confirming the utility of software hints. 
GRASP (Insertion-Only) \mbox{further} increases performance by yielding an average speed-up of 5.0\%. 
GRASP (Insertion-Only) provides additional \mbox{protection} to the High-Reuse cache blocks in comparison to RRIP+Hints by \mbox{inserting} High-Reuse cache blocks directly at the MRU position.
Finally, GRASP (Hit-Promotion) yields an average speed-up of 5.2\%. Difference between GRASP (Hit-Promotion) and GRASP (Insertion-Only) is marginal as the hit-promotion policy of GRASP has negative effect on slightly less than half the datapoints. 
The results are inline with the observations from prior work that showed that the value-addition of hit-promotion policies over insertion policies is low in presence of cache thrashing~\cite{dead}.

\noindentsectiontitle{\bf \em Summary:} Hardware cache management is an established difficult problem, which is reflected in the small average speed-ups (usually 1\%-5\%) achieved by state-of-the-art cache management schemes over the prior best schemes~\cite{rrip,sampler,ship,mdpp,hawkeye,perceptron,harmony}. Our work shows that graph applications present a particularly challenging workload for these schemes, in many cases leading to significant performance slowdowns. 
In this light, \sam is quite successful in improving performance of graph applications by yielding an average speed-up of 5.2\% (max 10.2\%) while not causing slowdown on any datapoint. Moreover, unlike state-of-the-art schemes, \sam achieves this without requiring storage-intensive~metadata.

\subsection{Pinning-based Schemes\label{sec:eval-pin}}

\begin{figure*}[!t]
    \centering
    \includegraphics[width=1\linewidth]{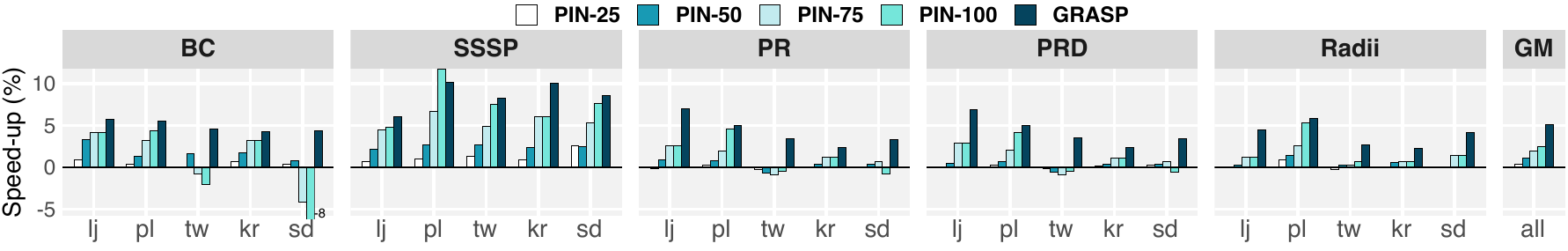}
    \reducefigcaptionspace
    \caption{Speed-up for \sam and pinning-based schemes over the RRIP baseline on high-skew~datasets.}
    \label{fig:sam-main2-perf}
    \vspace{-0.5em}
\end{figure*}
   
\begin{figure}[!t]
    \centering
    \includegraphics[width=1\linewidth]{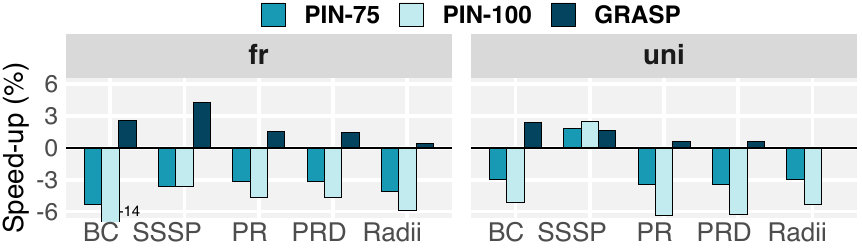}
    \reducefigcaptionspace
    \caption{Speed-up over the RRIP baseline on \fr{}, a low-skew dataset and \uni, a no-skew dataset.}
    \label{fig:sam-no-skew} 
\end{figure}

In this section, we show the benefit of flexible \sam policies over pinning-based rigid approaches. 
We first present the results on the high-skew datasets and then on the low-/no-skew datasets to test their resilience in adversarial scenarios.

\noindentsectiontitle{\bf \em High-skew datasets:} Fig.~\ref{fig:sam-main2-perf} shows speed-ups for four XMem configuration (PIN-25, PIN-50, PIN-75 and PIN-100) and GRASP over the RRIP baseline on high-skew datasets. 
\sam outperforms all XMem configurations on 24 of 25 datapoints with an average speed-up of 5.2\%. 
In comparison, PIN-25, PIN-50, PIN-75 and PIN-100 yield 0.4\%, 1.1\%, 2.0\% and 2.5\%, respectively.

PIN-100 outperforms the other three XMem configurations as for those configurations, a significant fraction of the capacity can still be occupied by cold vertices, which causes thrashing in the unreserved capacity.
Nevertheless, PIN-100 causes slowdown on many datapoints (\eg for BC, PR and PRD applications on \tw{} and \sd{} datasets). Moreover, PIN-100 cannot capitalize on reuse from Moderate Reuse Region as pinned vertices cannot be evicted even when they stopped exhibiting reuse.
Thus, PIN-100 provides only a marginal speed-up on many datapoints (\eg Radii application on \lj{}, \tw{} and \kr{} datasets).

PIN-75 and PIN-100, two best-performing XMem configurations, while yield only marginal speed-ups, still outperform the state-of-the-art history-based schemes, SHiP-MEM, Leeway and Hawkeye (Figs.~\ref{fig:sam-main1-perf} \& \ref{fig:sam-main2-perf}), which confirms that utilizing software knowledge for cache management is a promising direction over storage-intensive domain-agnostic design for the challenging access patterns of graph analytics.

\noindentsectiontitle{\bf \em Low-/No-skew datasets:} Next, we evaluate the robustness of \sam and pinning-based schemes (PIN-75 and PIN-100) for adversarial datasets with low-/no-skew. Naturally, these schemes are not expected to provide a significant speed-up in the absence of high skew; however, a robust scheme would reduce/avoid the slowdown. Fig.~\ref{fig:sam-no-skew} shows the speed-up for a low-skew dataset \fr{} and a no-skew dataset \uni{} for these schemes over the RRIP baseline.

\sam provides a net speed-up on 9 out of 10 datapoints even for low-/no-skew datasets. On the low-skew dataset \fr, \sam yields a speed-up between 0.4\% and 4.3\% whereas on the no-skew dataset \uni, \sam yields a speed-up between -0.1\% and 2.4\%. In contrast, PIN-75 and PIN-100 cause slowdown on almost all datapoints. 

In the absence of high skew, cache blocks belonging to the High Reuse Region do not dominate the overall LLC accesses. Thus, pinning these blocks throughout the execution is counter-productive for PIN-75 and PIN-100. In contrast, \sam adopts a flexible approach, wherein the high priority cache blocks from High Reuse Region can make way for other blocks that observe some reuse, as needed. Thus, \sam successfully limits slowdown, and even provides reasonable speed-up on some datapoints, for such highly adversarial datasets.

Finally, combining results on all 7 datasets (5 datasets from Fig.~\ref{fig:sam-main2-perf} and 2 from Fig.~\ref{fig:sam-no-skew}), GRASP yields an average speed-up of 4.1\%. In comparison, PIN-75 and PIN-100 provide a marginal speed-up of only 0.5\% and 0.1\%, respectively.
PIN-75 and PIN-100 cause slowdown of up to 5.3\% and 14.2\% whereas max slowdown for GRASP is only 0.1\%. 

\subsection{Reordering Techniques and GRASP}
Thus far, we evaluated GRASP on graph applications \mbox{processing} datasets that are reordered using DBG. In this section, we compare performance of vertex reordering techniques, followed by an evaluation of GRASP on top of these techniques, demonstrating GRASP's generality.

\noindentsectiontitle{\bf \em Effectiveness of Reordering Techniques: \label{eval-reordering-techniques}} In this section, we first show that skew-aware techniques can improve performance of graph applications even as a standalone software optimization, thus justifying their existence. We evaluate three skew-aware techniques -- Sort, HubSort~\cite{fc} and DBG~\cite{dbg} -- and a complex vertex reordering technique -- \gorder{}~\cite{gorder}.
We perform these studies on a real machine with 40 hardware threads as mentioned in Sec.~\ref{sec:sw-eval-method}. 

\begin{figure}[!t]
    \centering
    {\subfloat[\footnotesize Net speed-up for existing software reordering techniques after accounting for their reordering cost on a real machine.]{\includegraphics[width=1\linewidth]{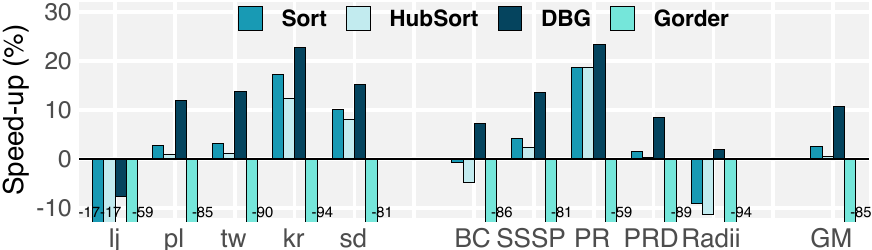}\label{fig:sw-perf}}}
    {\subfloat[\footnotesize Application speed-up of \sam over the RRIP baseline on top of different reordering techniques.]{\includegraphics[width=1\linewidth]{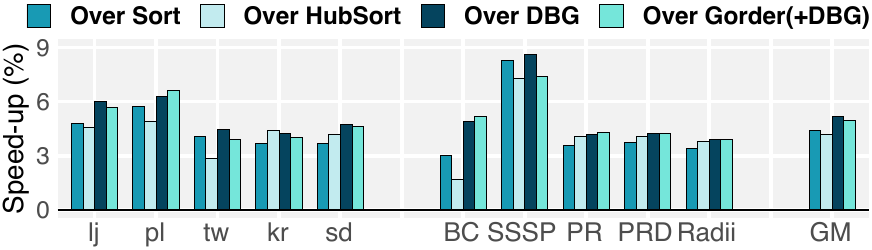}\label{fig:sam-compatibility}}}
    \caption{Reordering Techniques + GRASP: the left group shows speed-up for a dataset across all applications while the right group shows speed-up for an application across all~datasets.}
    \label{fig:graphin}
    \vspace{-0.5em}
\end{figure}

Fig.~\stitchref{fig:graphin}{fig:sw-perf} shows the speed-up for these existing software techniques after accounting for their reordering cost over the baseline with no reordering.
Among skew-aware techniques, all techniques are effective on largest of the datasets (\eg \kr{} and \sd) and long iterative applications (\eg \pr). As these techniques rely on a low cost approach for reordering, the reordering cost is amortized quickly when the application runtime is high, making these solutions practically attractive. Averaged across all application and dataset pairs, skew-aware techniques yield a net speed-up of 2.6\% for Sort, 0.6\% for HubSort and 10.8\% for DBG.

Unsurprisingly, \gorder{} causes significant slowdown on all datapoints due to its large reordering cost, yielding an \mbox{average} speed-up of -85.4\%. Thus, \gorder{} is less practical when \mbox{compared} to simple yet effective skew-aware techniques, corroborating prior work~\cite{dbg}.

\noindentsectiontitle{\bf \em Generality of \sam: \label{sam-compatibility}}
As software vertex reordering techniques offer different trade-offs in preserving graph structure and reducing reordering cost, it is important for \sam to not be coupled to any one software technique. In this section, we evaluate \sam with different reordering techniques, both skew-aware and complex ones. While skew-aware techniques are readily compatible with \samnospace, \gorder{} requires a simple tweak as follows.

After applying \gorder{} on an original dataset, we apply DBG to further reorder vertices, which results in a vertex order that retains most of the \gorder{} ordering while also segregating hot vertices in a contiguous region, making \gorder{} compatible with \samnospace. 

Fig.~\stitchref{fig:graphin}{fig:sam-compatibility} shows the speed-up for \sam over RRIP on top of the same reordering technique as the baseline.
As with DBG, \sam consistently provides a speed-up across datasets and applications on top of other reordering techniques as well. On average, \sam yields a speed-up of 4.4\%, 4.2\%, 5.2\% and 5.0\% on top of Sort, HubSort, DBG and \gorder, respectively. The result confirms that \sam complements a broad class of existing software reordering~techniques. 

\subsection{GRASP vs Optimal Replacement (OPT) \label{sec:oracle}}

In this section, we compare GRASP with Belady's optimal replacement policy (OPT)~\cite{opt}. As OPT requires the perfect knowledge of the future, we generate the traces of LLC accesses (up to 2 billion for each trace) for the applications processing graph datasets reordered using DBG on the simulation baseline configuration specified in Sec.~\ref{sec:hw-eval-method}.
We apply OPT on each trace for five different LLC sizes -- 1MB, 4MB, 8MB, 16MB and 32MB -- to obtain the minimum number of misses for a given cache size and report the percentage of misses eliminated over {\em LRU} on the same LLC~size.

\begin{figure}[!t]
    \centering
    \includegraphics[width=1\linewidth]{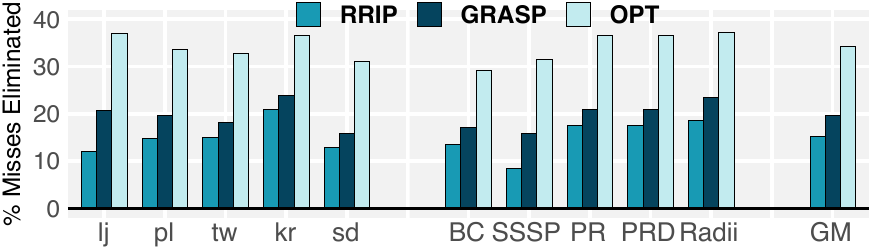}
    \reducefigcaptionspace
    \caption{Percentage of misses eliminated over LRU.}
    \label{fig:oracle}
    \vspace{-.4em}
\end{figure}

\noindentsectiontitle{\bf \em Miss reduction on 16MB LLC:}
Fig.~\ref{fig:oracle} shows the results for OPT along with RRIP and GRASP for 16MB LLC size.
OPT eliminates 34.3\% of total misses over LRU. In comparison, GRASP eliminates 19.7\% of misses (vs 15.2\% for RRIP). Overall, GRASP is 57.5\% effective in eliminating misses when compared to OPT, an offline technique with perfect knowledge of the future. While GRASP is the most effective among the online techniques, the results also show that the remaining opportunity (difference between OPT and GRASP) is still significant, which warrants further research in this~direction.

\noindentsectiontitle{\bf \em Sensitivity of GRASP to LLC size: }
Table~\ref{tab:size-sensitivity-oracle} shows the average percentage of misses eliminated by RRIP, \sam and OPT for different LLC sizes over LRU. With the increase in LLC size, \sam becomes more effective at eliminating misses over LRU (average miss reduction of 15.4\% for 1MB vs 21.2\% for 32MB). This is expected, as the larger LLC size allows \sam to provide preferential treatment to more hot vertices. In general, yet larger LLC sizes are expected to benefit even more from \sam until the LLC size becomes large enough to accommodate all hot vertices. 

\begin{table}[!t]
    \caption{\footnotesize{Percentage of misses eliminated over LRU for different LLC~size.}}
    \label{tab:size-sensitivity-oracle}
    \reducetablecaptionspace
    \centering
    \footnotesize
    \begin{tabular}{|l||c|c|c|c|c|c|}
        \hline
        \centering{\bf Scheme} & {\bf 1MB}	&  {\bf 4MB} & {\bf 8MB} & {\bf 16MB} & {\bf 32MB} \\ \hline \hline
        RRIP & 15.9\% &	16.4\% &	15.7\% &	15.2\% &	16.2\% \\ \hline
        GRASP & 15.4\% &	17.0\% &	18.1\% &	19.7\% &	21.2\% \\ \hline
        OPT	& 27.5\% &	32.2\% &	33.3\% &	34.3\% &	34.5\% \\ \hline
    \end{tabular}
\end{table}

\section{Related Work}
\label{related}

\noindent{\bf \em Shared-memory graph frameworks:}
A significant amount of research has focused on designing high performance shared-memory frameworks for graph applications. Majority of these frameworks are vertex-centric~\cite{ligra, galois, gap, graphmat, graphlab, graphchi} and use CSR or its variants to encode a graph, making GRASP readily compatible with these frameworks. More generally, GRASP requires classification of only the Property Array(s), making it independent of the specific data structure used to represent the graph, which further increases compatibility across the spectrum of frameworks. Thus, we expect GRASP to reduce misses across frameworks, though absolute speed-ups will likely vary.

\noindentsectiontitle{\bf \em Distributed-memory graph frameworks:}
these frameworks can also benefit from GRASP. For \mbox{example}, 
PGX~\cite{pgx.d} and PowerGraph~\cite{powergraph} proposed duplicating high degree vertices in the graph partitions to reduce high communication overhead across computing nodes. These optimizations are largely orthogonal to \sam cache management. As such, \sam can be applied to distributed graph processing by caching high-degree vertices within each nodes's LLC to improve node-level cache~behavior.

\noindentsectiontitle{\bf \em Streaming graph frameworks:}
In this work, we have \mbox{assumed} that graphs are static.
In practice, graphs may evolve over time and a stream of graph updates (\ie addition or removal of vertices or edges) are interleaved with graph-analytic queries (e.g., computing pagerank of vertices or computing shortest path from different root vertices). For such deployment settings, a CSR-based structure is infeasible. Instead, researchers have proposed various data structures for graph encoding that can accommodate fast graph updates and allow space-efficient versioning~\cite{stinger, llama, aspen}. Meanwhile, each graph query is performed on a consistent view (\ie static snapshot) of a graph. For example, Aspen~\cite{aspen}, a recent graph-streaming framework, uses Ligra (a static graph-processing framework) in the back-end to run graph-analytic queries. Thus, the observations made in this paper regarding cache thrashing due to the irregular access patterns of the Property Array, as well as skew-aware reordering and GRASP being complementary in combating cache thrashing, are also relevant for dynamic~graphs.

For static graphs, vertex reordering cost is amortized over multiple graph traversals for a single graph query (as shown in Fig.~\stitchref{fig:graphin}{fig:sw-perf}). However, for dynamic graphs, reordering cost can be further amortized over multiple graph queries.
Intuitively, addition or deletion of some vertices or edges in a large graph would not lead to a drastic change in the degree distribution, and thus unlikely to change which vertices are classified hot in a short time window. Therefore, skew-aware reordering can be applied at periodic intervals to improve cache behavior after a series of updates has been made to a graph, amortizing reordering cost over multiple graph queries.

\noindentsectiontitle{\bf \em Software-hints through profiling:}
Prior works have proposed ISA changes by embedding load/\mbox{store} instructions with reuse hints to improve cache replacement decisions~\cite{cache-hints1, cache-hints1-soft-assisted, cache-hints2, cache-hints3, cache-hints5-pacman}. These works perform program analysis via an additional compiler pass and/or runtime profiling to identify data that are unlikely to be referenced again. 
Compiler uses a custom memory instruction tagged with a low-reuse hint to access such data so that the cache hardware can prioritize the associated cache block for eviction.
Such mechanisms, however, are largely effective for only those programs that are dominated by loops with regular access patterns. For example, Pacman~\cite{cache-hints5-pacman} fails when a loop index variable and the reuse distance for the element accessed in a given loop iteration does not exhibit a linear correlation (e.g., indirect memory accesses, dominant type of memory accesses for graph analytics). In contrast, \sam leverages vertex placement in memory after skew-aware reordering pass to correctly learn high-reuse vertices, that too without requiring custom instructions or a priori program analysis.

\noindentsectiontitle{\bf \em Hardware prefetchers:} 
Modern processors typically employ prefetchers that target stride-based access patterns and thus are not amenable to graph analytics.  
Researchers have proposed custom prefetchers at L1-D that specifically target indirect memory access patterns of graph analytics~\cite{imp,prefetch-data-structure}. 
Nevertheless, prefetching can only {\em hide} memory access latency. Unlike cache replacement, prefetching cannot reduce memory bandwidth pressure or DRAM energy expenditure. 
Indeed, prior work observes that even a 100\% accurate prefetcher for graph analytics is bottlenecked by memory bandwidth~\cite{imp}. In contrast, GRASP reduces bandwidth pressure by reducing LLC misses, and thus is complementary to prefetching.

\noindentsectiontitle{\bf \em Traversal scheduling:}
Mukkara \etal proposed HATS~\cite{hats}, a hardware-accelerator implementing locality-aware scheduling to exploit cache locality for graphs exhibiting community structure. While effective, it requires intrusive hardware changes, including a  specialized hardware unit with each core and an ISA change on the host core. In contrast, \sam requires a minimal hardware interface and trivial changes to the cache policy while largely utilizing the existing cache~hardware.

\noindentsectiontitle{\bf \em Graph slicing:}
Researchers have proposed slicing, a software optimization that slices the graph in LLC-size partitions and processes one partition at a time to reduce irregular memory accesses.
Specifically, Graphicionado~\cite{graphicionado} uses slicing to fit a working set into a large 64MB on-chip scratchpad while Zhang \etal use \emph{CSR Segmenting} to break the vertices into segments that fit in LLC~\cite{fc}.

While generally effective, slicing has two important limitations. First, it requires invasive framework changes to form the slices (which may include replicating vertices to avoid out-of-slice accesses) and manage them at runtime. Secondly, for a given cache size, the number of slices increases with the size of the graph, resulting in greater processing overheads in creating and maintaining partitions for larger graphs.

In comparison, \sam is a hardware scheme that intelligently leverages lightweight software support. \sam requires minimal changes in a graph framework and does not require any changes in the graph algorithms.
Having said that, \sam is complementary to slicing; by preserving the critical working set in the cache (i.e., hot vertices), \sam could be used to improve the performance of Graphicionado to reduce the number of slices by making intra-slice cache misses infrequent. 

\section{Conclusion}
\label{conclusion}
{\nohyphens
This work explores how hardware cache management should be designed to tackle cache thrashing at LLC for graph-analytic applications. We show that state-of-the-art cache management schemes are deficient in presence of cache thrashing stemming from irregular access patterns of graph applications processing large graphs. In response, we introduce \samnospace~-- specialized cache management for LLC for graph analytics on power-law graphs. \samnospace's specialized cache policies exploit the high reuse inherent in hot vertices while retaining the flexibility to capture reuse in other cache blocks. \sam leverages existing software reordering optimizations to enable a lightweight interface that allows hardware to pinpoint hot vertices amidst irregular access patterns. In doing so, \sam avoids the need for a storage-intensive prediction mechanism or additional metadata storage in the LLC. \sam requires minimal hardware support, making it attractive for integration into a commodity server processor to enable acceleration for the domain of graph analytics. Finally, GRASP delivers consistent performance gains on high-skew datasets, while preventing slowdowns on low-skew datasets.
}

\section*{Acknowledgment}
{\nohyphens
This work was supported in part by a research grant from Oracle Labs. We thank Amna Shahab, \mbox{Antonios} \mbox{Katsarakis}, Arpit Joshi, Artemiy \mbox{Margaritov}, Avadh Patel, Dmitrii Ustiugov, Rakesh Kumar, Vasilis Gavrielatos, Vijay Nagarajan, and the anonymous \mbox{reviewers} for their valuable feedback on an earlier draft of this work. 
}

{
\footnotesize
\nohyphens
\printbibliography
}

\end{document}